%% file: main.tex
\documentclass[aps,pre,twocolumn,showpacs,superscriptaddress,10p,bibliography]{revtex4-2}

\usepackage{amsmath}
\usepackage{amssymb}
\usepackage[sort&compress]{natbib}
\usepackage{graphicx}
\usepackage{color}
\usepackage{physics}
\usepackage{appendix}

\graphicspath{{../Images/}}

\usepackage{float}
\usepackage[caption = false]{subfig}
\begin{document}

\title{Quantum dynamics of a $ 4\pi$-kink in Josephson junctions parallel arrays with large kinetic inductances}
\author{S. S. Seidov}
\affiliation{National University of Science and Technology ``MISIS", Moscow 119049, Russia}
\author{M. V. Fistul}
\affiliation{Theoretische Physik III, Ruhr-Universit\"at Bochum, Bochum 44801, Germany}

\begin{abstract} 
We present a theoretical study of the quantum dynamics of  \textit{two} magnetic fluxons (MFs) trapped in Josephson junction parallel arrays (JJPAs) with large kinetic inductances. The Josephson phase distribution of two trapped MFs satisfies a topological constraint, i.e., a total variation of Josephson phases along a JJPA is $4\pi$. In such JJPAs the characteristic length of Josephson phase distribution ("the size" of MF) is drastically reduced to be less than a single cell size. 
Two extreme dynamic patterns will be distinguished: two weakly interacting MFs and two merged MFs, i.e., a $4\pi$--kink.
Taking into account the repulsive interaction between two MFs located in the same or adjacent cells we obtain the energy band spectrum $E_{4\pi}(p)$ for a quantum $4\pi$--kink. 
The coherent quantum dynamics of a $4\pi$--kink demonstrates the quantum beats with the frequency and amplitude strongly deviating from ones observed for two independent MFs. In the presence of applied dc and ac bias current of frequency $f$ a weakly incoherent quantum dynamics of a $4\pi$-kink results in the  Bloch oscillations and the seminal current steps with values $I^{(n)}_{4\pi}=enf$ which are two times less than ones for two independent MFs.
\end{abstract}

\maketitle
\section{Introduction}
Magnetic fluxons (MFs) are topological solitons \cite{manton2004topological,kartashov2011solitons,scott1999nonlinear,braun2004frenkel} studied intensively in various low--dimensional superconducting systems, e.g., in Josephson junction parallel arrays (JJPAs) \cite{kivshar1989dynamics,ustinov1998solitons,mazo2014sine,Carapella}. Each MF is a vortex of superconducting current carrying one magnetic flux quantum, $\Phi_0$.
The JJPAs composed of a large number of superconducting cells with embedded Josephson junctions, have provided a well established experimental platform for studying of the MFs dynamics. Variety of fascinating physical phenomena in the classical nonlinear dynamics of MFs have been observed experimentally, e.g.,  dc/ac current induced resonances \cite{kivshar1989dynamics}, the relativistic dynamics of MF \cite{ustinov1998solitons}, bunching of MFs \cite{vernik1996soliton}, the Cherenkov radiation of plasma modes by moving MF \cite{wallraff2000whispering}, ac current induced dynamic metastable states \cite{fistul2000libration} and the dynamics of MFs in a specially prepared ratchet potential \cite{ustinov2004ratchetlike}. The MF itself is a $2\pi$--kink in the spatial distribution of Josephson phases \cite{kivshar1989dynamics,ustinov1998solitons}. The effective methods have been elaborated to trap, manipulate and to measure MFs in JJPAs. 

Even more complex macroscopic objects, e.g. the breathers, i.e. (anti)MF-MF pairs, observed in long Josephson junctions \cite{ustinov1998solitons,mazo2014sine}, discrete breathers trapped in Josephson junction ladders \cite{binder2000observation,trias2000discrete} or high-order kink states in JJPAs \cite{vernik1996soliton,peyrard1984kink,champneys2000origin,ustinov1998bunched,pfeiffer2006observation}, have been theoretically and experimentally studied. The latter have been predicted long time ago \cite{peyrard1984kink,champneys2000origin} and, in spite of the repulsion of two MFs approaching to each other,  moving \textit{$4\pi$--kinks} have been experimentally observed in Ref. \cite{pfeiffer2006observation}. The fingerprint of the $4\pi$--kink nonlinear classical dynamics is a specific branch of the current-voltage characteristics ($I$-$V$ curve) substantially deviating from the one related to the motion of two independent MFs. Since the classical dynamics of MFs and high-order kinks in JJPAs is governed by a large set of coupled nonlinear differential equations with topological constraints \cite{ustinov1998solitons,mazo2014sine} the effective numerical procedures have been developed for a theoretical study of that. 

A new twist in this field, namely a study of the \textit{quantum} dynamics of macroscopic topological objects, has attracted a great interest. Initial studies have demonstrated a large number of incoherent quantum phenomena such as the macroscopic quantum tunneling of a bunch of MFs in two-dimensional Josephson junction arrays \cite{fazio2001quantum,van1996quantum}, the macroscopic quantum tunneling and energy level quantization of a single MF \cite{kato1996macroscopic,hermon1994quantum,shnirman1997tunneling,wallraff2000annular,wallraff2003quantum}, and the quantum dissociation of a vortex-antivortex pair \cite{fistul2003quantum} in long annular Josephson junctions.  It was realized that
 the main obstacle on the way to observe the \textit{coherent} quantum-mechanical effects in the dynamics of MFs is a large spatial extent (the MFs "size") of a $2\pi$--kink greatly exceeding the size of a single cell. Indeed, in such JJPAs various non-topological macroscopic objects, e.g. the plasma oscillations or vortex-antivortex pairs, can easily be excited leading to an additional dissipation and decoherence in the dynamics of MFs. In order to drastically decrease the MFs size one needs to replace low geometrical inductances by large kinetic ones.  Implementation of large kinetic inductances in JJPAs or long Josephson junctions can be provided by two effective methods:  an embedding of series arrays of large Josephson junctions in each cell of JJPAs \cite{manucharyan2009fluxonium,moskalenko2021quantum,Moskalenko}, or using disordered superconducting materials \cite{maleeva2018circuit,hazard2019nanowire,astafiev2012coherent,wildermuth2022fluxons}.

The coherent quantum dynamics of a single MF trapped in a JJPA with large kinetic inductances have been theoretically studied in \cite{seidov2021quantum,petrescu2018fluxon,moskalenko2021quantum}. In such JJAs the  Josephson phases  strongly vary from a one cell to adjacent cells, and therefore, almost anti-continuous limit is established \cite{braun2004frenkel,joos1982properties,seidov2021quantum}. Such coherent quantum-mechanical effects as the quantum beats of a single MF in JJAs composed of a few cells \cite{moskalenko2021quantum}, the MF energy band \cite{seidov2021quantum,petrescu2018fluxon}, complex quantum oscillations controlled by the Aharonov-Casher phase, weakly incoherent dynamics of quantum MF leading to the macroscopic Bloch oscillations \cite{seidov2021quantum} have been studied in detail. Therefore, it is a good starting point to theoretically study the coherent quantum dynamics of high-order kinks trapped in JJAs. 

In this Article we extend elaborated previously by us \cite{seidov2021quantum} analysis of the quantum dynamics of a single MF to the case of \textit{two} MFs trapped in a JJPA with high kinetic inductances. The Josephson phase distribution of two MFs monotonically increases along the JJPA, and the total variation of  Josephson phases is equal to $4\pi$. In the anticontinuous limit each MF is characterized by Josephson phases of three consecutive Josephson junctions \cite{seidov2021quantum}, and using this approximation  we derive the repulsive interaction potential of two MFs and present the detailed study of the coherent quantum dynamics of two interacting MFs. In the analysis we distinguish two extreme dynamical patterns: the quantum dynamics of two independent MFs and a $4\pi$-kink, i.e., two merged MFs. For both cases we obtain the energy bands determining the coherent motion of two MFs along the JJPA, the time-dependent probability to find both MFs in a fixed cell of the JJPA, and macroscopic Bloch oscillations occurring in the presence of a weak dissipation. A  quantitative comparison of these dynamic patterns allows one to obtain distinguished features of the $4\pi$-kink coherent quantum dynamics.   
 

The paper is organized as follows: In Section II we present our model for  JJPAs with large kinetic inductances, provide the generic expression for the potential energy $U(\{\varphi_i\})$, where $\{\varphi_i\}$ are the Josephson phases of individual  Josephson junctions.  In Sec. III we study the effective potential energy of two magnetic fluxons trapped in a  JJPA. For that we use a special approximation where a single fluxon is characterized by Josephson phases of three consecutive Josephson junctions \cite{seidov2021quantum}. In Section IV we elaborate a two-dimensional tight-binding model allowing to study the coherent quantum dynamics of two interacting magnetic fluxons. The quantum-mechanical dispersion relation of a $4\pi$--kink will be obtained. In Sections V and VI we discuss the specific quantum properties of a $4\pi$-kink and compare that with two independent MFs. The Section VII provides conclusions. 

\section{JJPAs with large kinetic inductances: model and potential energy}
\begin{figure}[h!!]
\subfloat[a)]{\includegraphics[width = \linewidth]{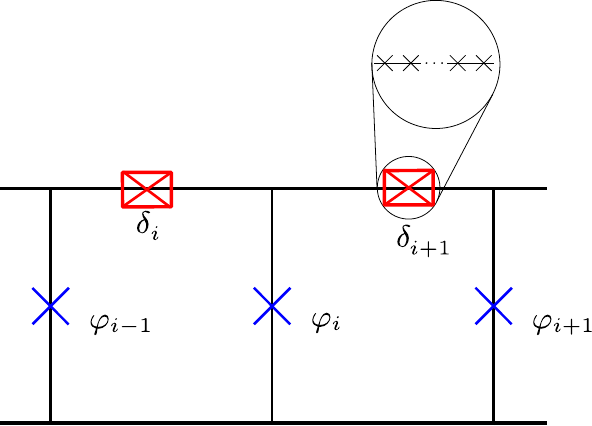}}\\
\subfloat[b)]{\includegraphics[width = \linewidth]{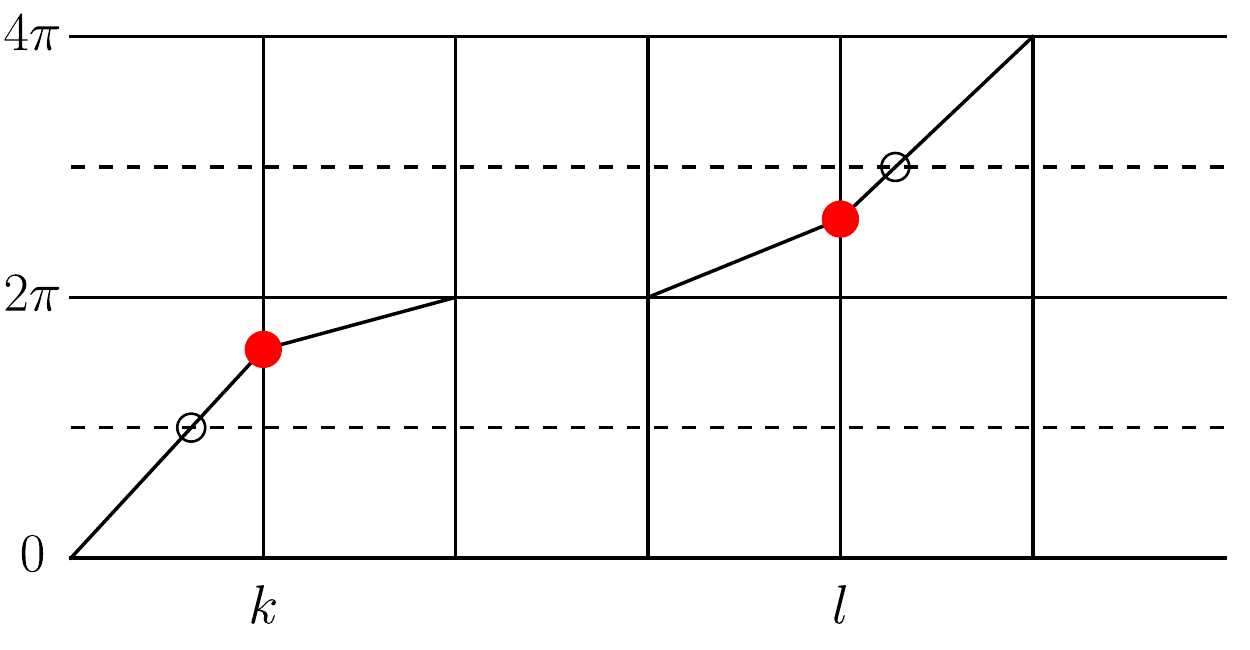}}
\caption{a) Schematic of a JJPA with large kinetic inductances. Small Josephson junctions with Josephson phases $\varphi_i$ are shown by crosses. Large kinetic inductances are provided by series arrays of large Josephson junctions (shown by red boxes). 
b) The typical Josephson phase distribution of two well separated MFs trapped in a JJPA. 
Red dots represent values of Josephson phases in the centres of the MFs, open circles --- "positions" of the MFs.}
\label{fig:System}
\end{figure}

Let us consider a JJPA composed of $M$ small (quantum) Josephson junctions incorporated in superconducting cells of large kinetic inductances. The cell size is $d$. The dynamics of small Josephson junctions is determined by time-dependent Josephson phases, $\varphi_i(t)$, and these Josephson junctions can demonstrate the quantum-mechanical behavior on a macroscopic scale. Small Josephson junctions are characterized by two important physical parameters: the Josephson coupling energy, $E_J$ and the charging energy, $E_c$. Large kinetic inductances of superconducting cells are provided by series arrays of large (classical) Josephson junctions, $\delta_i$. Moreover, these series arrays of Josephson junctions allow to effectively block the undesirable penetration of magnetic fluxes into JJAs. An external magnetic field piercing JJPA cells is characterized by the magnetic fluxes, $\Phi_i$.
The schematic of such setup is shown in Fig. \ref{fig:System}a. 

A general expression for the JJPA potential energy with the Josephson phase distribution $\{\varphi_i \}$ is given by 
\begin{equation}\label{eq:U}
\begin{aligned}
&U\qty(\{\varphi_i\}) = E_J \sum_{i=1}^{M} (1 - \cos\varphi_i) + \\
&+ E_L \sum_{i=1}^{M} \qty(\varphi_i - \varphi_{i+1} + 2\pi \frac{(n_i + \Phi_i)}{\Phi_0})^2.
\end{aligned}
\end{equation}
Here $E_L$ is the kinetic inductance energy  that is supposed to be small in respect to the Josephson coupling energy of individual Josephson junctions, $E_J$, i.e., $E_L \ll E_J$. In next 
we consider a simplest case as all $\Phi_i, n_i$ are set to zero.

\section{JJPAs with two trapped magnetic fluxons} 
Next, we consider a particular case as two MFs are trapped in the JJPA. A single MF bearing the magnetic flux quantum $\Phi_0$ is the $2\pi$-kink in the Josephson phase distribution $\{\varphi_i \}$, and correspondingly, in the presence of two MFs the $\{\varphi_i\}$ has to be a monotonic function and to satisfy the following topological constraint: the total variation of $\varphi_i$ along the JJPA is $4\pi$. 

To obtain the potential energy of two interacting MFs trapped in the JJPA with large kinetic inductances ($E_J \gg E_L$) we use the method elaborated previously to study the classical \cite{joos1982properties} and quantum \cite{seidov2021quantum} dynamics of a single MF in the anti-continuous limit, where a single MF is characterized by three consecutive Josephson phases, and other Josephson phases are set to $0$ or $2\pi$. To apply this method for the JJPA with two trapped MFs we fix the centers of MFs in cells $k$ and $\ell$. The Josephson phases of MFs centers are $\varphi_k$ varying in the region between $0$ and $2\pi$, and $\varphi_l$ varying in the region between $2\pi$ and $4\pi$, accordingly. The typical Josephson phase distribution of two well separated MFs is presented in Fig. \ref{fig:System}b. 

\subsection{Well separated MFs: $l - k \geq 3$}
As the centers of MFs are located on the large distance, e.g., $l-k = 3$ (see the Josephson phase distribution in Fig. \ref{fig:System}b), the Josephson phase distribution is written in the following form:
\begin{equation}\label{Phasedistr-3}
\begin{aligned}
\{\varphi_i\} = \{&0, \ldots, 0, \varphi_{k-1}, \varphi_k, 2\pi + \tilde \varphi_{k+1}, 2\pi + \tilde \varphi_{l-1}, \\
& 2\pi+\tilde \varphi_l, 4\pi-(2\pi-\tilde \varphi_{l+1}), 4\pi, \ldots, 4\pi\},
\end{aligned}
\end{equation} 
where we introduce the renormalized Josephson phase as $\tilde \varphi_n = \varphi_n - 2\pi$. Substituting  (\ref{Phasedistr-3}) in (\ref{eq:U}) and taking into account that the Josephson phases $\varphi_{k-1}$, $|\tilde \varphi_{k+1}|$, $\tilde \varphi_{l-1}$ and $|\tilde \varphi_{l+1}-2\pi|$ are small, we expand  the potential energy (\ref{eq:U}) up to the second order with respect to these phases and minimize. Following this procedure and taking into account the terms up to the second order in $E_L/E_J$, we obtain the effective potential energy of two \textit{noninteracting} MFs:
\begin{equation}\label{eq:U3}
\begin{aligned}
&U_{\text{eff}}(\varphi_k, \tilde \varphi_{l = k + 3}) = 2E_L\qty(1 - \frac{2 E_L}{E_J})\qty(\varphi_k^2 + \tilde \varphi_l^2) -\\
&- 4 \pi E_L \qty(1 - \frac{2 E_L}{E_J})\qty(\varphi_k + \tilde \varphi_l) +\\
&+ E_J (2 - \cos\varphi_k - \cos \tilde \varphi_l) + U_0, 
\end{aligned}
\end{equation}
where $U_0$ is obtained explicitly as
\begin{equation}
U_0 = 8 E_L \pi^2 - 16\pi^2 \frac{E_L^2}{E_J}.
\end{equation}
Thus, one can see that the minimum of the effective potential energy occurs for small values of $\varphi_k$ and $\tilde \varphi_l $, and  the minimal value of (\ref{eq:U3}) is $E_0 = U_0 - 16\pi^2 E_L^2/E_J$. Notice here that in the limit of $E_J \gg E_L$ the Eq. (\ref{eq:U3}) is valid also for MFs located at the distance, $l-k \geq 3$. 

\subsection{Two MFs located at the distance: $l - k = 2$}
\begin{figure}[h!!]
\subfloat[a)]{\includegraphics[width = 0.49\linewidth]{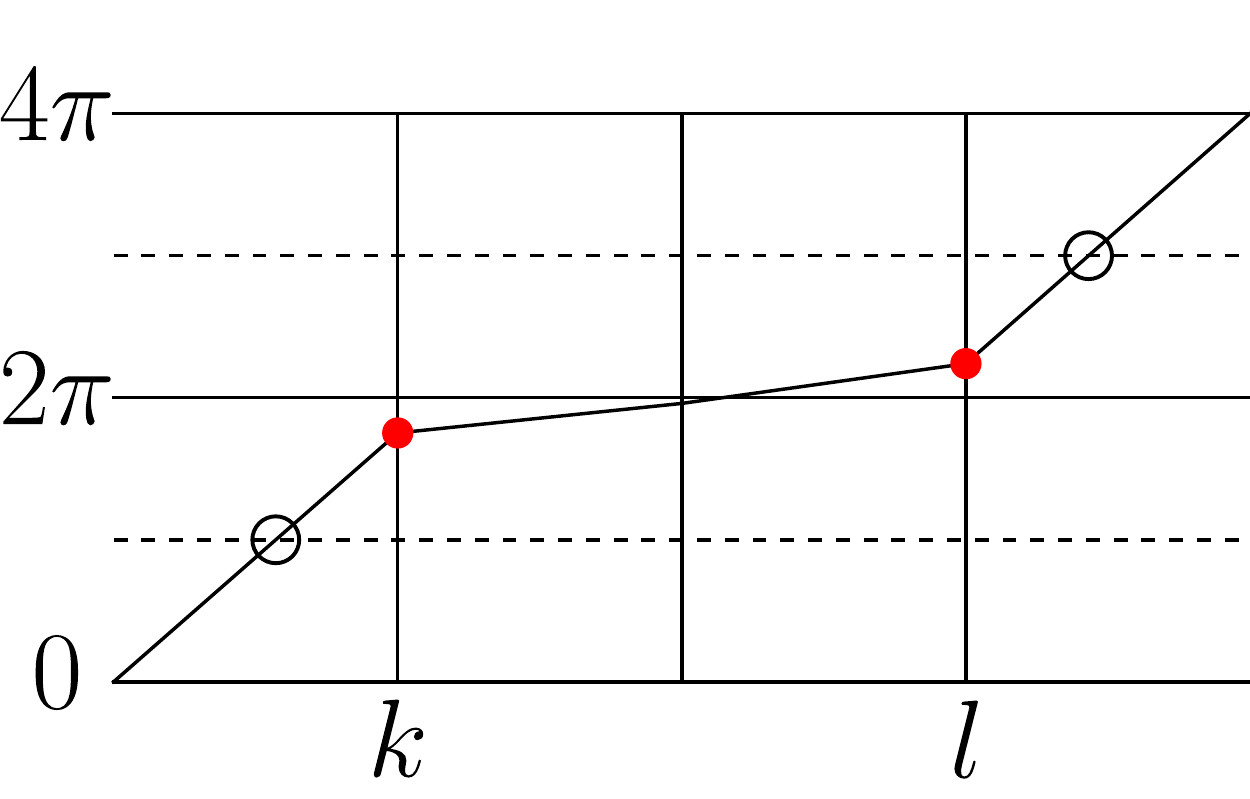}}
\subfloat[b)]{\includegraphics[width = 0.49\linewidth]{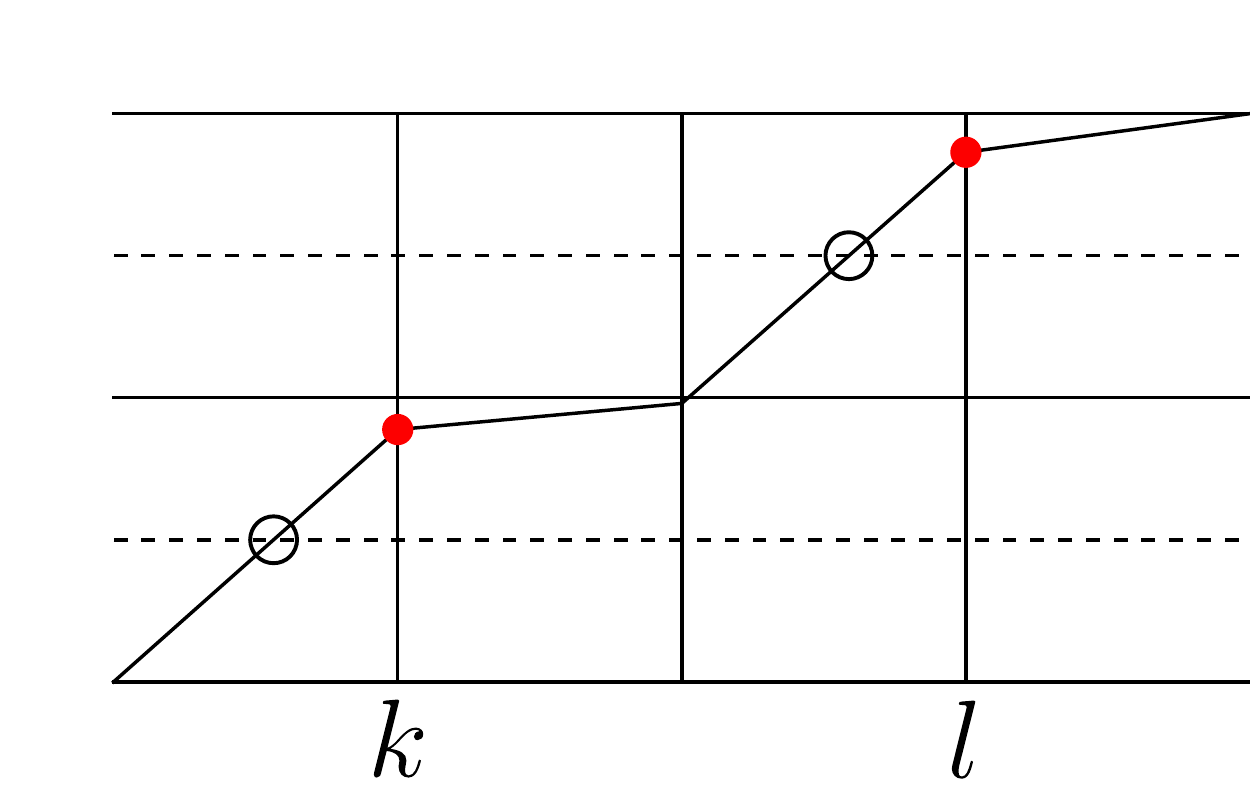}}\\
\subfloat[c)]{\includegraphics[width = 0.49\linewidth]{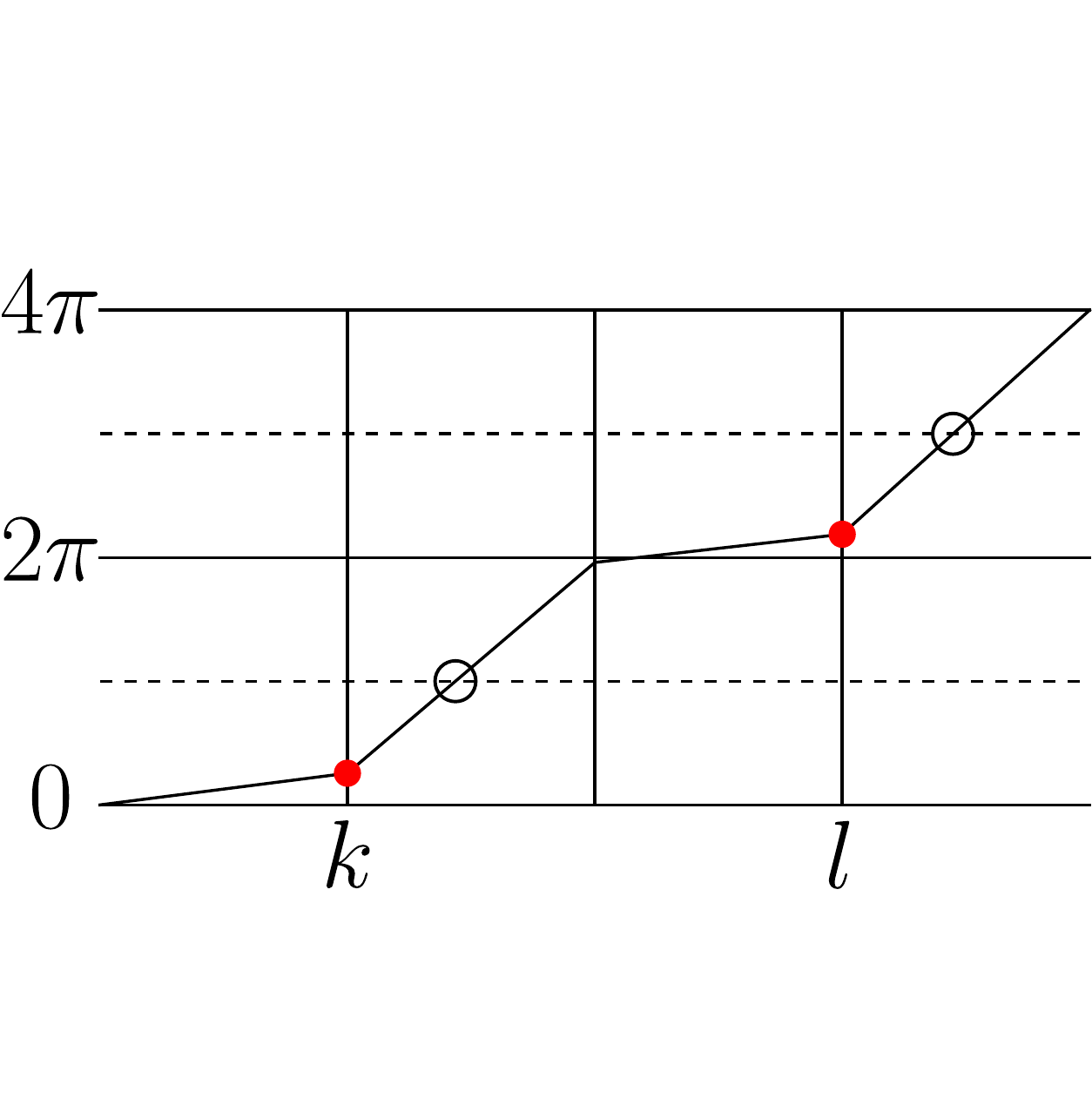}}\hspace{0.001\linewidth}
\subfloat[d)]{\includegraphics[width = 0.49\linewidth]{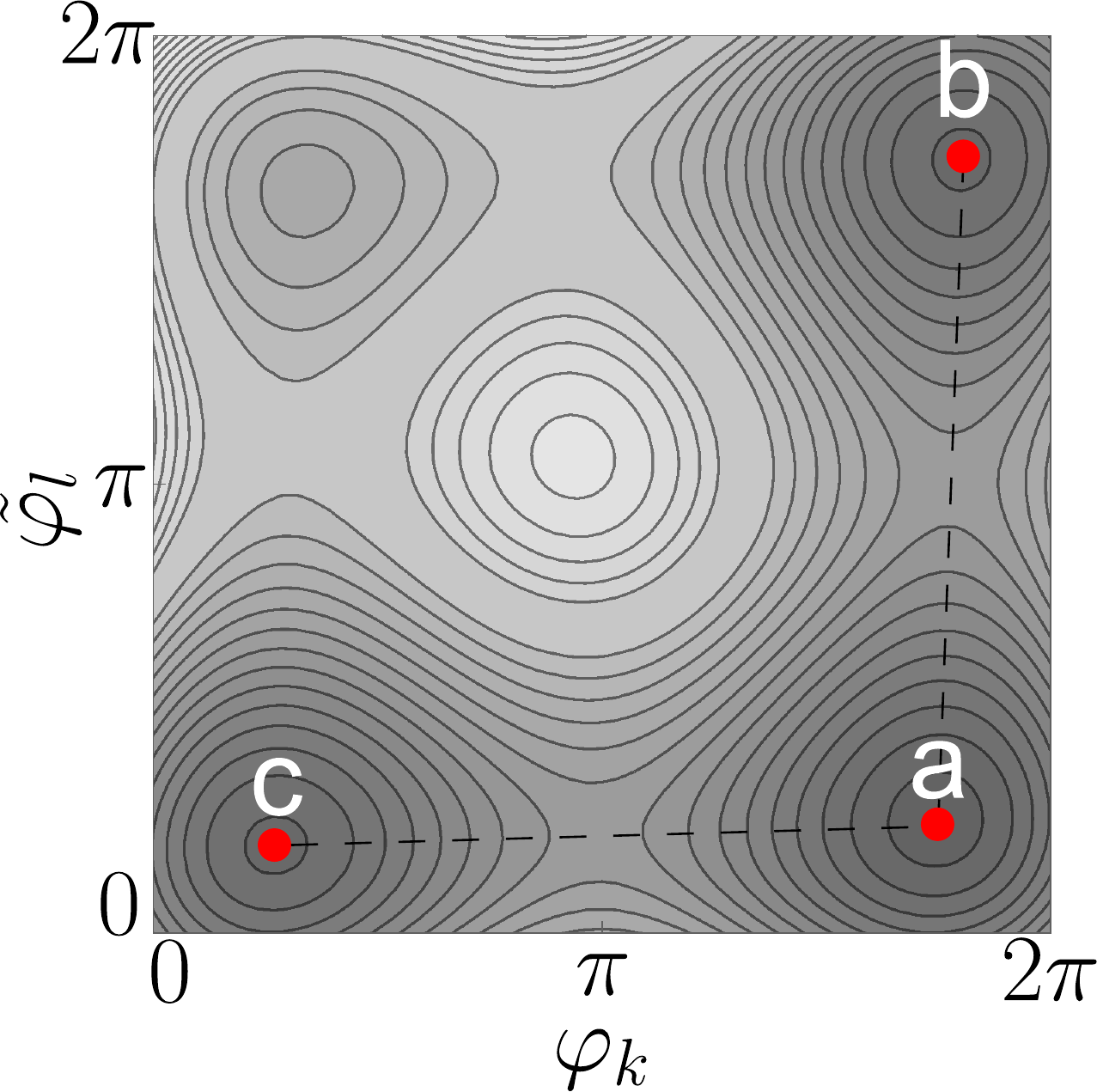}}
\caption{The configurations of Josephson phases (a), b), c)) in the minimums of $U_{\text{eff}}(l-k=2)$, and the contour plot (d)) of the dependence of the effective potential energy on the Josephson phases, $\varphi_k$ and $\tilde \varphi_l$, i.e., Eq. (\ref{eq:U2}). The parameters were chosen as $E_J = 1$, $E_L = 0.1$.}
\label{fig:U2}
\end{figure}

Now we bring the MFs closer and the configuration of Josephson phases is
\begin{equation}
\begin{aligned}
\{\varphi_i\} = \{&0, \ldots, 0, \varphi_{k-1}, \varphi_k, 2\pi + \tilde \varphi_{k+1}, \varphi_l,\\
&4\pi -(2\pi-\tilde \varphi_{l+1}),4\pi, \ldots, 4\pi\}.
\end{aligned}
\end{equation} 
By making use of the procedure analogous to the previous subsection we obtain the effective potential energy as
\begin{equation}\label{eq:U2}
\begin{aligned}
&U_\text{eff}(\varphi_k, \tilde \varphi_{l = k + 2}) = 2 E_L\qty(1 - \frac{2 E_L}{E_J})\qty(\varphi_k^2 + \tilde \varphi_l^2) -\\
&- 4 \pi E_L \qty(1 - \frac{2 E_L}{E_J})\qty(\varphi_k + \tilde \varphi_l) -\\
&- \frac{4 E_L^2}{E_J} (2\pi - \varphi_k)\tilde\varphi_l + E_J (2 - \cos\varphi_k - \\
&- \cos \tilde \varphi_l) + U_0.
\end{aligned}
\end{equation}
In the limit of $E_L \ll E_J$ the effective potential energy has three minimums located near the points: $\{\varphi_k, \tilde \varphi_l\} = \{0, 0\}, \{2\pi, 0\}, \{2\pi, 2\pi\}$. Minimizing the Eq. (\ref{eq:U2}) over the Josephson phases $\varphi_k$ and $\tilde \varphi_l$ we obtain explicit locations of the minimums (we mark it with letters $a$, $b$ and $c$ in the Fig. \ref{fig:U2}) and the minimal energy, $U_\text{min}$. 
The potential energy $U_\text{min}$ at each minimum is the same up to the second order in $E_L/E_J$:
\begin{equation}
\begin{aligned}
&U_\text{min}(|l-k|=2) = 8 E_L \pi^2 - 32 \pi^2 \frac{E_L^2}{E_J} 
&= E_0.
\end{aligned}
\end{equation}
Therefore, two MFs located at the distance $|l-k|=2$ do not interact with each other. 
In Fig. \ref{fig:U2} the configurations of the Josephson phases in minimums of $U_\text{eff}(l-k=2)$ and the contour plot of the effective potential energy  are presented.

In the quantum-mechanical regime the MFs can "hop" between the minimums due to the macroscopic tunneling of the Josephson phases. Thus, the hopping amplitude is determined by the potential energy profile between points $a$ and $b$ in the direction of $\varphi_k = \text{const}$ (see the dashed line in the contour plot of Fig. \ref{fig:U2}). We stress here that this profile is identical to the Josephson phase dependence of the potential energy of a single MF, i.e. the $2\pi$-kink, $U_{\text{eff}}^{2\pi}$ obtained in \cite{seidov2021quantum} as
\begin{equation}\label{eq:U2pi}
\begin{aligned}
&U_{\text{eff}}^{2\pi}(\tilde\varphi_l) = 2 E_L \qty(1 - \frac{2 E_L}{E_J}) \tilde\varphi_l^2 - \\
&- 4 \pi E_L \qty(1 - \frac{2E_L}{E_J})\varphi_l + E_J (1 - \cos \tilde\varphi_l) + \operatorname{const}.
\end{aligned}
\end{equation}

\subsection{Two MFs located at the distance: $l - k = 1$}

\begin{figure}[h!!]
\subfloat[a)]{\includegraphics[width = 0.49\linewidth]{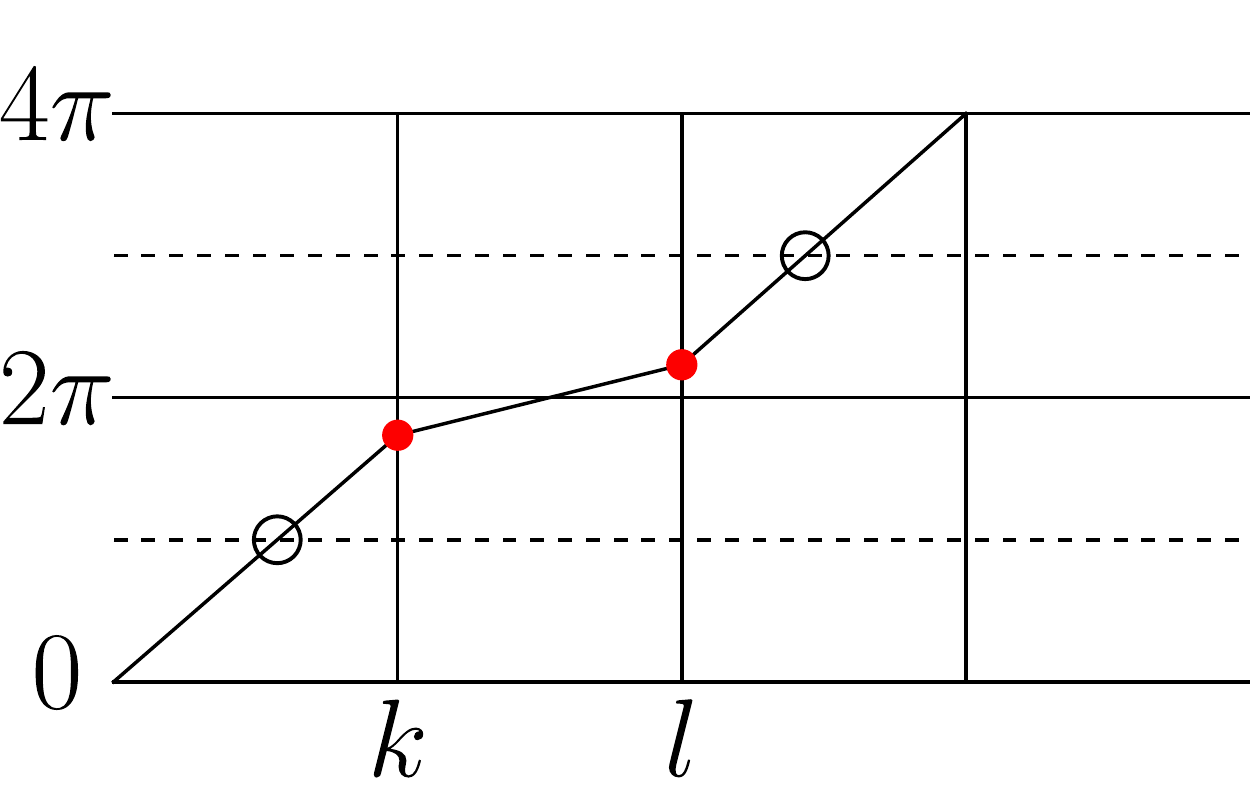}}
\subfloat[b)]{\includegraphics[width = 0.49\linewidth]{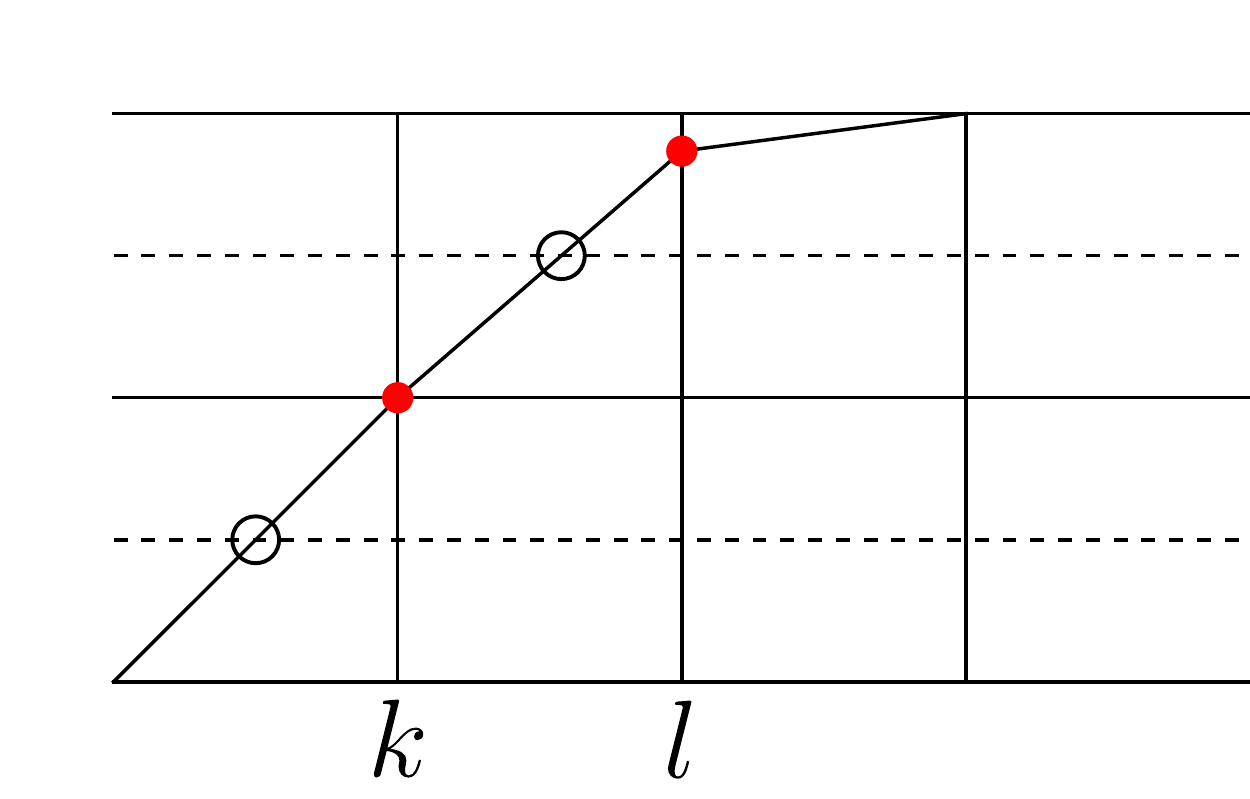}}\\
\subfloat[c)]{\includegraphics[width = 0.49\linewidth]{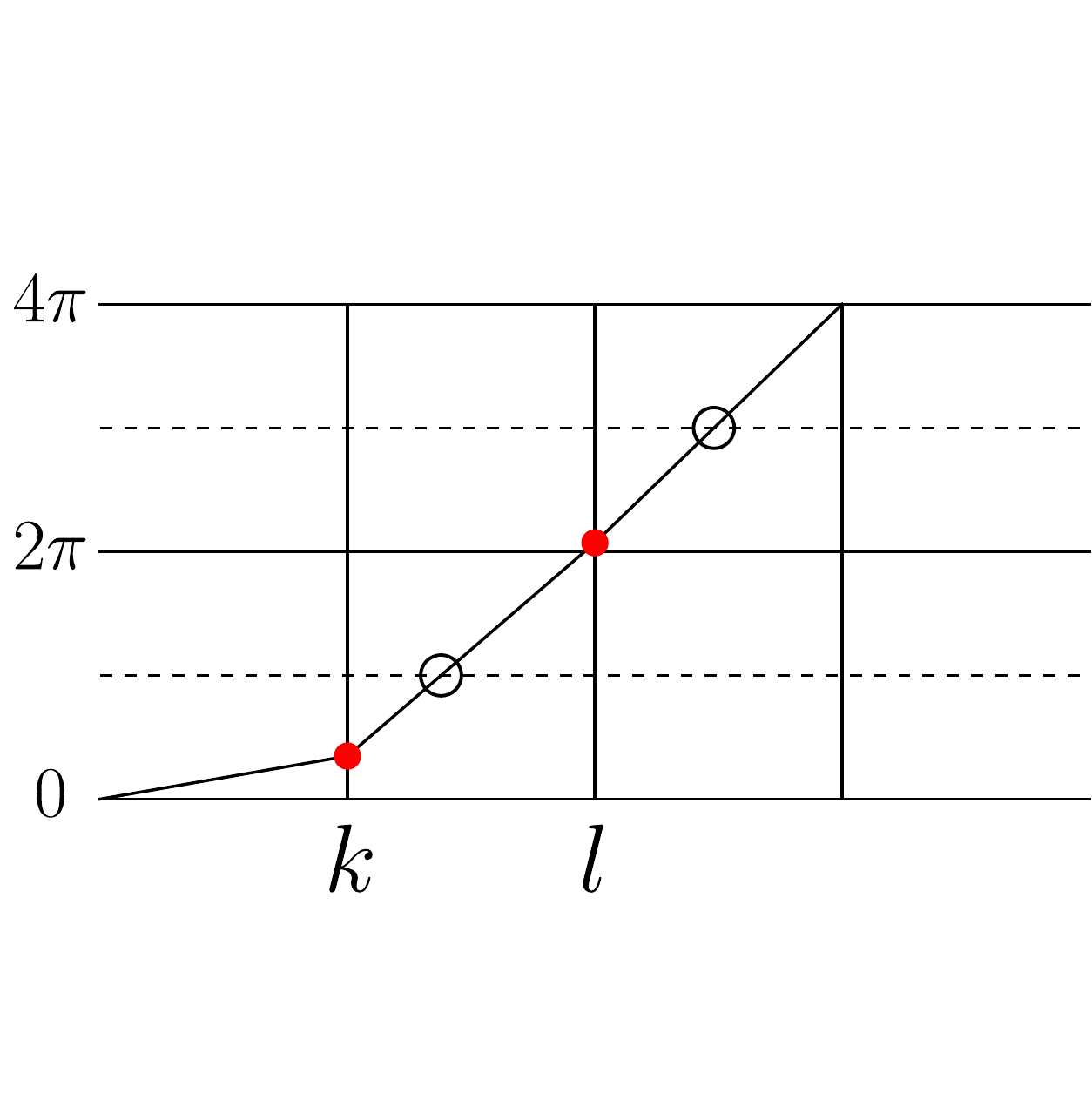}}\hspace{0.001\linewidth}
\subfloat[d)]{\includegraphics[width = 0.49\linewidth]{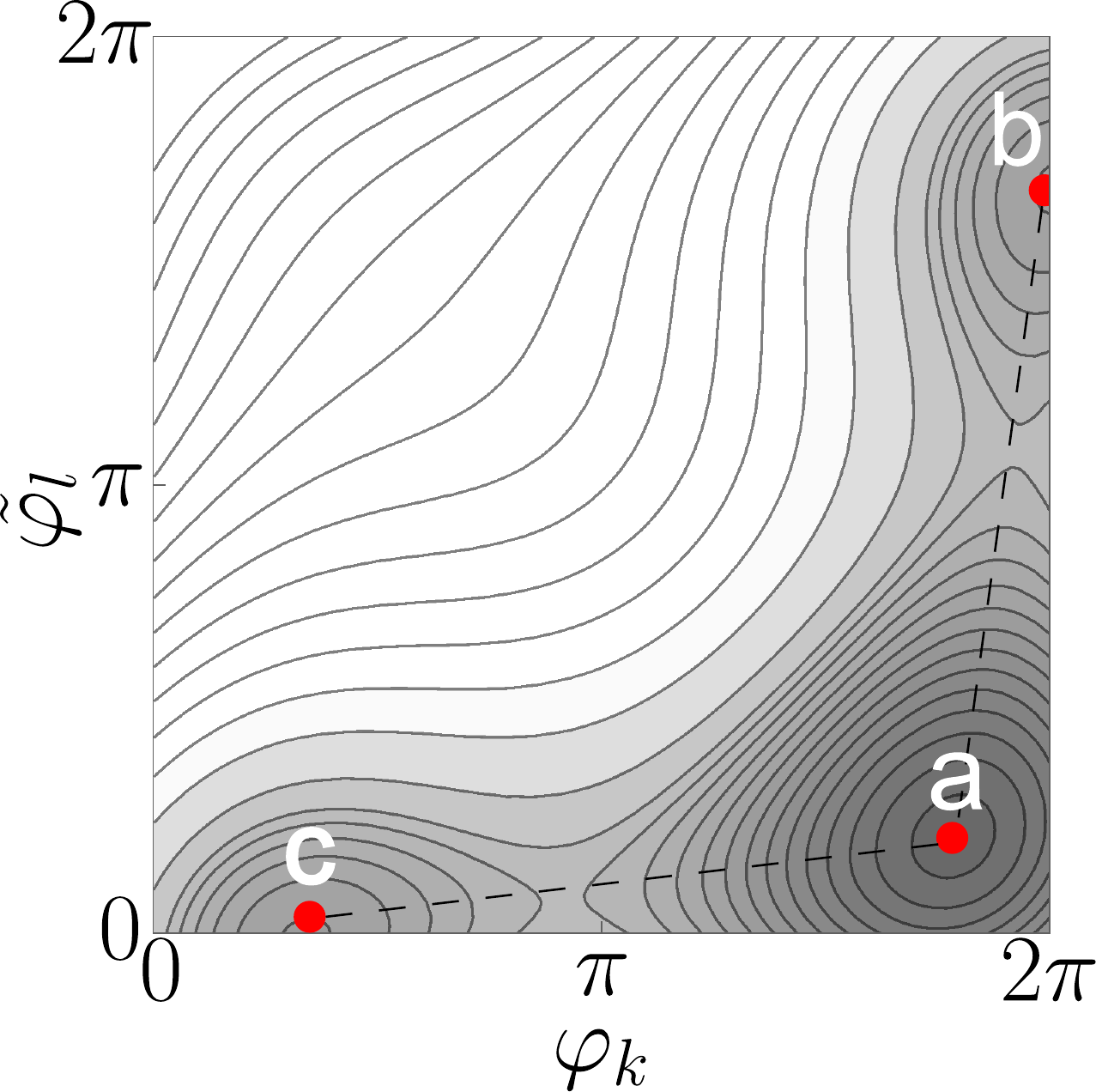}}
\caption{The configurations of Josephson phases (a), b), c)) in the minimums of $U_\text{eff}(l-k=1)$, and the contour plot (d)) of the dependence of the effective potential energy on the Josephson phases, $\varphi_k$ and $\tilde \varphi_l$, i.e., Eq. (\ref{eq:U1}). The parameters were chosen as $E_J = 1$, $E_L = 0.1$.}
\label{fig:U1}
\end{figure}

As the centers of two MFs are located in adjacent cells the Josephson phase configuration is written as 
\begin{equation}
\begin{aligned}
\{\varphi_i\} = \{&0, \ldots, 0, \varphi_{k-1}, \varphi_k, \varphi_l, 4\pi - (2\pi-\tilde \varphi_{l + 1}), \\
&4\pi, \ldots, 4\pi\}.
\end{aligned}
\end{equation} 
Using the procedure elaborated in subsections $A$ and $B$ we obtain the effective potential energy as  
\begin{equation}\label{eq:U1}
\begin{aligned}
&U_\text{eff}(\varphi_k, \tilde \varphi_{l = k + 1}) = 2 E_L\qty(1 - \frac{2 E_L}{E_J})\qty(\varphi_k^2 + \tilde \varphi_l^2) -\\
&- 4 \pi E_L \qty(1 - \frac{E_L}{E_J})\qty(\varphi_k + \tilde \varphi_l) +\\
&+\frac{4E_L^2}{E_J}\pi (\tilde \varphi_l - \varphi_k ) + 2 E_L (2\pi - \varphi_k) \tilde\varphi_l +\\
&+ E_J (2 - \cos\varphi_k - \cos \tilde \varphi_l) + U_0 - 8\pi^2 \frac{E_L^2}{E_J}.
\end{aligned}
\end{equation}
The expression (\ref{eq:U1}) demonstrates the interaction between two MFs. Since $2\pi - \varphi_k \geqslant 0$ the interaction term is a positive one, and the MFs repel from each other. 
The effective potential energy has three minimums marked as $a$, $b$ and $c$ in Fig. \ref{fig:U1}, and the values of $U_\text{eff}(l-k=1)$ in these minimums are given by 
\begin{equation}
\begin{aligned}
&U_\text{min} (|l-k|=1)\eval_{a} \approx 8 E_L \pi^2 - 32 \pi^2 \frac{E_L^2}{E_J} = E_0\\
&U_\text{min} (|l-k|=1)\eval_{b,c} \approx 8 E_L \pi^2 - 16 \pi^2 \frac{E_L^2}{E_J} = U_0 = E_1.
\end{aligned}
\end{equation}
Here we denote the energy of higher minimum as $E_1$. 

The tunneling amplitude from the state $a$ to the state $b$ is determined by the potential energy profile in $a-b$ direction that in the limit of $E_J \gg E_L$ is equal to $U_{\text{eff}}^{2\pi}(\varphi)$ (see Eq. (\ref{eq:U2pi})).

\subsection{Two MFs located in the same cell: $4\pi$-kink}
\begin{figure}[h!!]
\center\includegraphics[width = \linewidth]{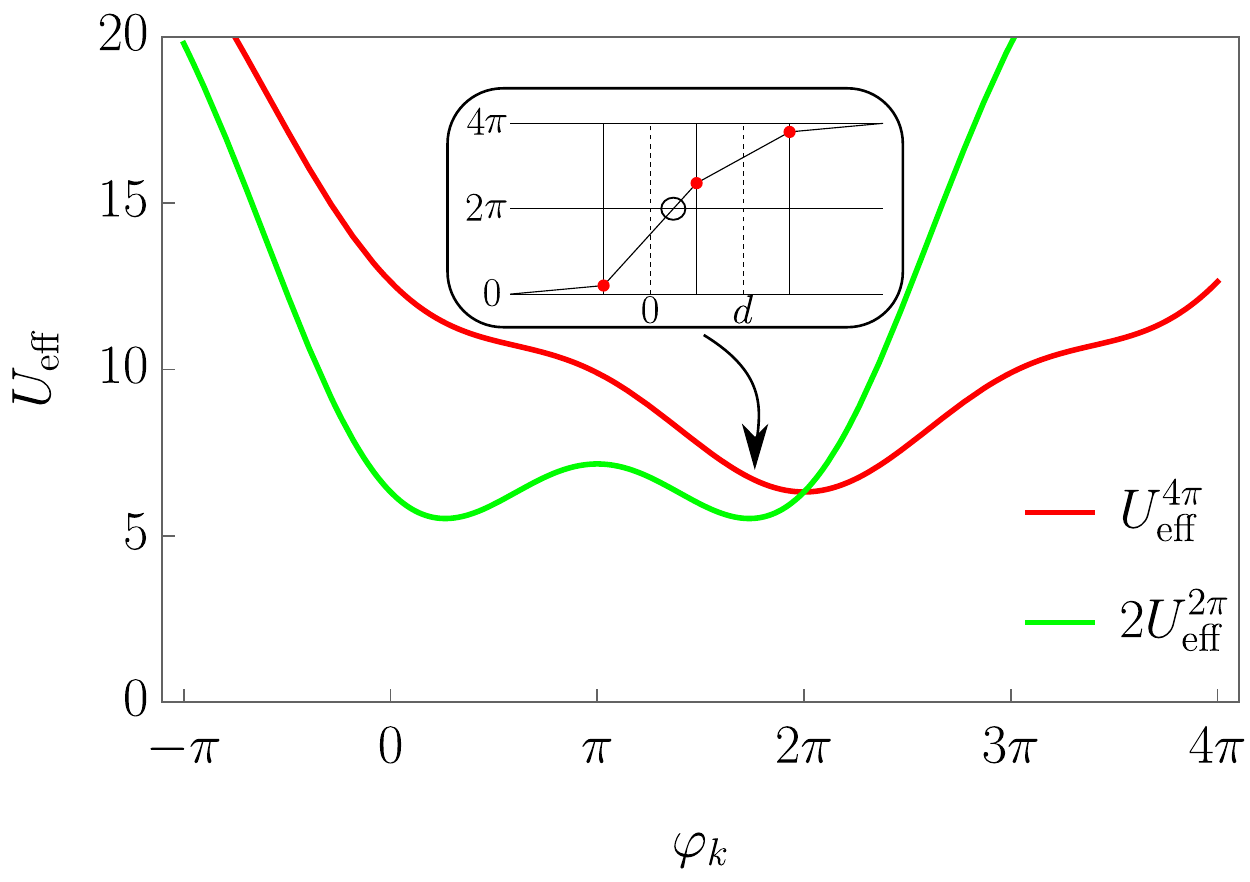}
\caption{Dependencies of the effective potential energies on the Josephson phase in the center of $4\pi$-kink (red line) and two independent $2\pi$-kinks located on the same position (green line).  The Josephson phase configuration of a stable $4\pi$-kink is shown in the inset. 
The parameters were chosen as $E_J = 1$, $E_L = 0.1$.}
\label{fig:U0}
\end{figure}

Here, we consider two MFs located in the same cell, i.e.,  $4\pi$--kink:
\begin{equation}
\begin{aligned}
\{\varphi_i\} =& \{0, \ldots, 0, \varphi_{k-1}, \varphi_k,\\
 &4\pi - (2\pi- \tilde \varphi_{k + 1}), 4\pi, \ldots, 4\pi\}.
\end{aligned}
\end{equation}
For this Josephson phase configuration the potential energy is
\begin{equation}
\begin{aligned}
&U_\text{eff}^{4\pi}(\varphi_k) = 2 E_L\qty(1 - \frac{2 E_L}{E_J})\varphi_k^2 -\\
&- 8 \pi E_L \qty(1 - \frac{2E_L}{E_J})\varphi_k + E_J (1 - \cos\varphi_k) + 2 U_0.
\end{aligned}
\end{equation}
The dependence of $U_\text{eff}^{4\pi}(\varphi_k)$ on the Josephson phase $\varphi_k$ in the center of $4\pi$--kink, is shown in Fig. \ref{fig:U0}. For comparison, the potential energy of two independent MFs located in the same cell, i.e. two $2\pi$--kinks,  is also presented in Fig. \ref{fig:U0}. 

The effective potential energy $U_{\text{eff}}^{4\pi}(\varphi_k)$ has a global minimum exactly at $\varphi_k = 2\pi$ and its value at the minimum is $E_1=U_0$. 
As one can see the $4\pi$--kink potential energy minimum is slightly higher than the potential energy of two independent MFs, and it coincides with the potential energy of two MFs in the configurations $b,c$ presented in Fig. \ref{fig:U1}.  

\section{Energy spectrum of two interacting MFs: tight-binding model with interaction}
\begin{figure}[h!!] 
\center\includegraphics[width=\linewidth]{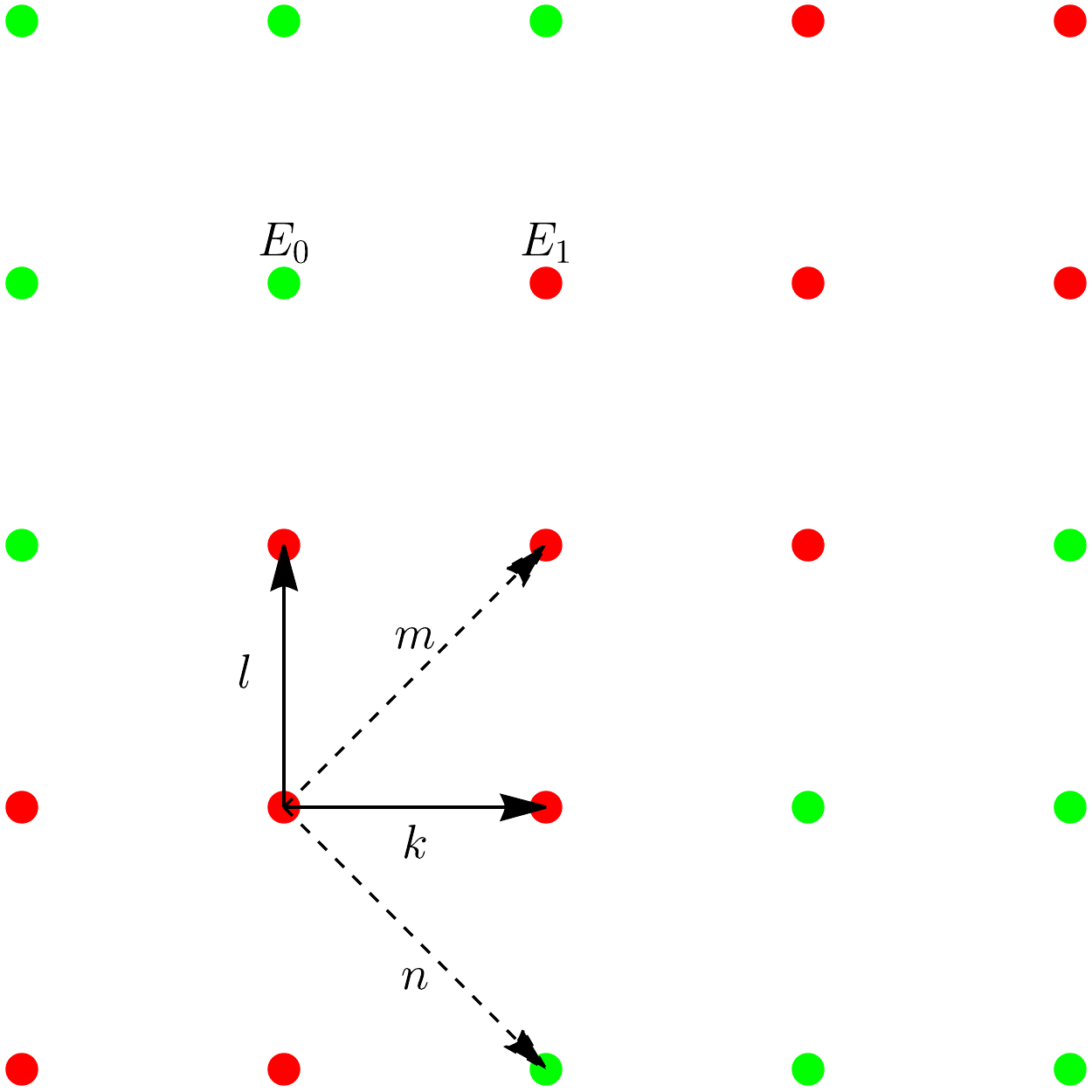}
\caption{The two-dimensional grid for the tight-binding model of two interacting quantum MFs. Such tight-binding model has a defect extended along $k = l, l \pm 1$ directions (indicated by red dots).  Red dots have the higher energy $E_1$ and green dotes have the lower energy $E_0$. Vectors $\vec{k}$, $\vec{l}$ and $\vec{m}=\vec{k}+\vec{l}$ and $\vec{n}=\vec{k}-\vec{l}$ are shown.}
\label{fig:grid}
\end{figure}

To quantitatively analyze the macroscopic quantum-mechanical phenomena we derive the energy spectrum of two interacting MFs. Taking into account the kinetic energy of two MFs expressed as, $K\{\dot \varphi_k,\dot \varphi_l \}=E_J/(2\omega_p^2)[\dot \varphi_k^2+\dot \varphi_l^2]$, where the plasma frequency $\omega_p=\sqrt{8E_JE_c}{\hbar}$ was introduced, and the dependence of the potential energy $U_\text{eff}$ on the distance $|l-k|$ between the cells $k$ and $l$, where the centers of two MFs are located, we arrive on the tight-binding model for two interacting \textit{quantum} MFs trapped in the JJPA. The quantum dynamics of two interacting MFs is determined by the macroscopic wave function $\ket{\Psi}$ that is the superposition of localized wave functions defined on the two-dimensional grid, i.e.
\begin{equation} \label{eq:psi}
\ket{\Psi} = \sum_{kl} c_{k,l} \ket{k,l}.
\end{equation}
Here, positive integers $k$ and $l$ indicate the numbers of cells where the centers of MFs are located; $c_{k,l}$ are the quantum-mechanical amplitudes. The localized states energies are $E_0$ if $|l - k| > 1$ and $E_1$ for $l=k$ and $l=k \pm 1$. The tunneling between the minimums of the  potential $U_{\text{eff}}$ provides the hopping amplitude $\Delta$ between the localized states on adjacent grids. Notice here, that the tunneling is absent between the states with the energy $E_1$, i.e., $k=l$ and $l=k \pm 1$. Summarizing that we obtain the tight-binding Hamiltonian in the following explicit form
\begin{equation}
\label{Hamiltonian}
\begin{aligned}
&\hat H = \sum_{kl} E_{kl} \ketbra{k, l} -\\
&- \frac{\Delta}{2}{\sum_{kl}}^\prime  \Big(\ketbra{k,l}{k + 1, l} + \ketbra{k,l}{k, l + 1} + \operatorname{h.c.} \Big)\\
&E_{kl} = E_0 + (E_1 - E_0)(\delta_{k,l} + \delta_{k, l + 1} + \delta_{k, l - 1}),
\end{aligned}
\end{equation} 
where the parameter $\Delta \simeq (\hbar \omega_p) \exp[-8E_J/(\hbar \omega_p)]$  has been calculated in \cite{seidov2021quantum,petrescu2018fluxon,moskalenko2021quantum}, and ${\sum}^\prime$ indicates the absence of tunneling between the grids $k=l$ and $l=k \pm 1$.
The schematic of such tight-binding model is presented in Fig. \ref{fig:grid}. This procedure resembles a one elaborated previously for the analysis of two interacting quantum particles (e.g. bosons) moving on a one-dimensional periodic lattice \cite{ivanchenko2014quantum,yusipov2017quantum}.

For an explicit calculation of the energy spectrum of two interacting MFs 
it is convenient to align one of the coordinate axis in $k=l$ direction and to use other integers $n = k - l$ and $m = k + l$ with the corresponding modification of the grids, $(k,l) \rightarrow (m,n)$. In this representation we rewrite the Hamiltonian as
\begin{equation}\label{eq:mn}
\begin{aligned}
& \hat H = \sum_{\substack{kl \\ m=k+l \\ n=k-l}} E_{mn} \ketbra{m, n} -\\
&- \frac{\Delta}{2} {\sum_{\substack{kl \\ m=k+l \\ n=k-l}}}^\prime\Big(\ketbra{m,n}{m + 1, n + 1} + \\
&+\ketbra{m,n}{m + 1, n - 1} + \operatorname{h.c.} \Big),\\
&E_{mn} = E_0 + (E_1 - E_0)(\delta_{n,0} + \delta_{n, 1} + \delta_{n,- 1})
\end{aligned}
\end{equation}
and the wave function has a following form: $\ket\psi = \sum_{k,l} c_{m,n} \ket{m,n}$. 
To obtain the energy spectrum $E$ and wave functions $\psi$ of two interacting quantum MFs 
the stationary Schr\"odinger equation for amplitudes $c_{m,n}$ has to be written. The stationary Schr\"odinger equation presents a set of difference equations with the lattice defect extended along  $n = 0, \pm 1$ directions:
 the energy of cites $n=0, \pm 1$, is $E_1 > E_0$. We treat cites with indices $n = -1, 0, 1$ as a single cite because they correspond to the same MFs configuration. Thus, the direct tunneling from the cites $n = -1, 0, 1$ to the cites $n = \pm 2$ is allowed. The difference equations for coefficients $c_{m,n}$ on a whole grid are
\begin{equation} \label{eq:Schr2}
\begin{aligned}
&(E_0 - E)c_{m,n} -  \frac{\Delta}{2} (c_{m+1,n+1} + c_{m+1,n-1} +\\
&+c_{m-1,n+1} + c_{m-1,n-1}) = 0,\ |n| > 1;\\
&(E_1 - E)c_{m,1} - \frac{\Delta}{2} (c_{m+1,2} + c_{m+1,-2} + \\
&+c_{m-1,2} + c_{m-1,-2}) = 0; \\
& c_{m,-1}=c_{m,0}=c_{m,1}.
\end{aligned}
\end{equation}

\subsection{Scattering states of two interacting quantum MFs}
First, we describe the \textit{scattering} states of two interacting MFs. Far away from the defect ($|n| >1  $) we search the solution $c_{m,n}$ for such states in the form of plain waves 
\begin{equation}\label{eq:scattering}
\begin{aligned}
c_{m,n}&= \exp \left \{i\frac{p_1 kd + p_2 l d}{\hbar} \right\}=\\
&= \exp \left \{i\frac{p (md/2) + q n d}{\hbar} \right \},
\end{aligned}
\end{equation}
where $p_1$ and $p_2$ are quasi-momenta of first and second MFs. The $p = (p_1+p_2)$ and $q = (p_1-p_2)/2$ are the center of mass and relative quasi-momenta of two  MFs, accordingly. 

Substituting this expression in a first equation of (\ref{eq:Schr2}) we obtain the two-dimensional energy band spectrum $E(p,q)$ as  
\begin{equation}\label{eq:MFS-2pi}
E_{2 \times 2\pi} (p, q)= E_0 - 2\Delta \cos \left [ \frac{pd}{2\hbar }\right ]\cos \left [ \frac{qd}{\hbar }\right].
\end{equation}
Thus, one can see that the quantum dynamics of two MFs with the energy spectrum (\ref{eq:MFS-2pi}) is determined by a weak scattering as the centers of two MFs approach to each other on the distance of $d$. The energy band spectrum $E_{2 \times 2\pi}(p,q)$ is just a sum of energies of independent MFs.

\subsection{The bound states of two interacting MFs: $4\pi$-kink quantum dynamics}
Beyond the scattering states considered in the subsection $A$ the Hamiltonian
(\ref{eq:mn}) supports the \textit{bound states } solution. The bound states amplitudes $c_{m,n}$ are caught in the following form:
\begin{equation}\label{eq:Boundstates}
c_{m,n} = e^{i p m d/(2\hbar)} [-\operatorname{sign} (\eta)]^n e^{- |n| d/\lambda},
\end{equation}
where $\lambda$ is the characteristic length of the bound state of two MFs, and $\eta=\cos [pd/(2\hbar)]$
Substituting (\ref{eq:Boundstates}) in the first equation of (\ref{eq:Schr2}) we obtain the energy $E_{4\pi}$ as following: 
\begin{equation}\label{eq:MFs-4pi}
E_{4\pi} = E_0 + 2\Delta |\cos [pd/(2\hbar)]|\cosh (d/\lambda)
\end{equation}
Substituting (\ref{eq:MFs-4pi}) in the second equation of (\ref{eq:Schr2})  we obtain $\lambda$, and finally, the energy spectrum $E_{4\pi} (p)$ as
\begin{equation}\label{eq:Epm}
E_{4\pi}(p) = E_0 + \sqrt{4 \Delta^2 \cos^2[p d/(2\hbar)] + (E_1 -E_0)^2}.
\end{equation}
Thus, one can see that the bound states of two MFs are characterized by a one-dimensional energy band spectrum, $E_{4\pi}(p)$ determining the quantum dynamics of a $4\pi$-kink trapped in the JJPA. 

\section{The coherent quantum oscillations of two interacting MFs}
\begin{figure}[h!!]
\center\includegraphics[width = \linewidth]{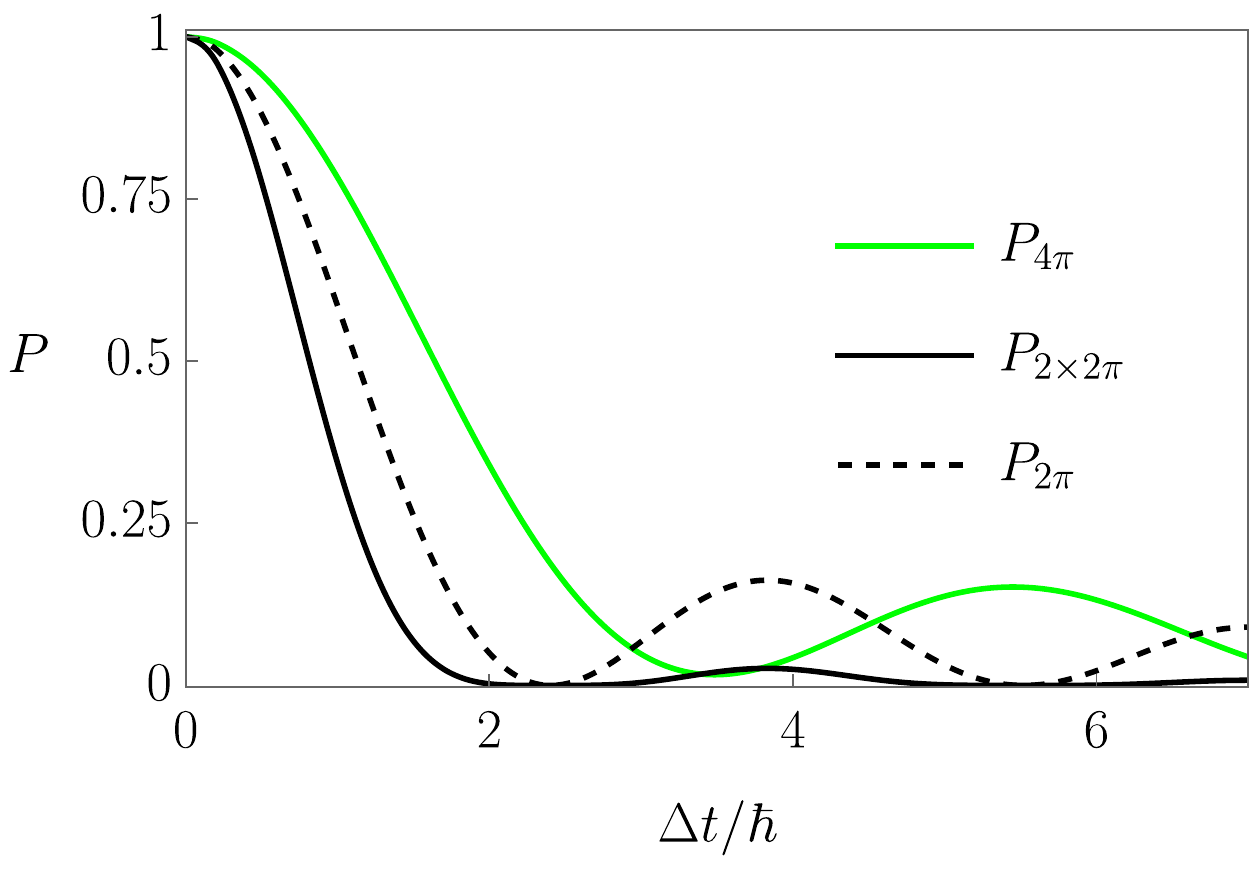}
\caption{Probabilities of finding $4\pi$-kink (green solid line), two independent fluxons (black solid line) and a single fluxon ( black dashed line) in cell number $0$ at time $t$. Here, we chose the parameter $\Delta/(E_0 - E_1) = 1/\sqrt{2}$.}
\label{fig:P}
\end{figure}

In the previous Section we obtain that the quantum dynamics of two MFs trapped in a one-dimensional JJPA can be realized in two forms: two weakly interacting moving MFs (the scattering states) or moving $4\pi$-kink (the bound states). The coherent quantum dynamics of two MFs is determined by the probability $P(k,l;t)$ to find the MFs in cells $k$ and $l$ at the time $t$ if initially both MFs were in the cell $0$. For $4\pi$-kink we obtain
\begin{equation}\label{eq:P}
\begin{aligned}
P_{4\pi}(k=l,t) &= \left |d \int\limits_{-\pi \hbar/d}^{\pi \hbar/d} \frac{\dd p}{2\pi \hbar}\exp \left\{ -\frac{i E_{4\pi}(p) t}{\hbar} - \right. \right.\\
&-\left. \left.  \frac{ip k d}{\hbar} \right\} \right|^2,
\end{aligned}
\end{equation}
Taking into account the energy band spectrum of  $4\pi$-kink, i.e., (\ref{eq:Epm}), one can calculate numerically $P_{4\pi}(k=l,t)$ for different values of $\Delta$ and $(E_1-E_0)$. The typical time dependence of $P_{4\pi}(0,t)=P_{4\pi}(t)$, i.e. the \textit{return} probability, is presented in Fig. \ref{fig:P}. Moreover, in the limit of $(E_1-E_0) \gg \Delta$ one can obtain an explicit expression: $P_{4\pi}(t)=J_0^2 [\frac{\Delta^2}{(E_1-E_0)\hbar}t]$, where $J_0(x)$ is the Bessel function \cite{Abramowitz}.
Similarly, for scattering states of two weakly interacting MFs we obtain the $P_{2\times 2\pi }(k,l;t)$ as following 
\begin{equation}
\begin{aligned}
&P_{2\times 2\pi }(k,l;t) = \left |d^2 \iint \frac{\dd p\dd q}{(2\pi \hbar)^2} \exp \left\{-\frac{2i\Delta}{\hbar} t \times   \right. \right.\\
&\left. \left. \times \cos \frac{p d}{2\hbar} \cos \frac{q d}{\hbar} - \frac{iq (k-l) d }{\hbar} - \frac{ip (k+l) d}{2\hbar} \right\} \right|^2.
\end{aligned}
\end{equation}
The probability $P_{2\times 2\pi }(0,0;t)=P_{2\times 2\pi }(t)$ is calculated explicitly as $P_{2\times 2\pi }(t)=J_0^4 [\Delta t/\hbar]$ \cite{Abramowitz}. Both time dependencies of $P_{4\pi}(t)$ and $P_{2 \times 2\pi}(t)$ are presented in Fig. \ref{fig:P}. 
For comparison the time-dependence of the return probability for a single MF, $P_{2\pi}(0,t)=P_{2\pi}(t)=J^2_0 (\Delta t/\hbar)$, is also shown in Fig. \ref{fig:P}.  

To conclude this section we notice that the applied gate voltage $V_g$ provides the macroscopic Aharonov-Cashier phase $\chi \propto V_g$ allowing one to  vary the return probability $P_{4\pi}(t)$ in short annular JJPAs \cite{seidov2021quantum}.

\section{Weakly dissipative quantum dynamics of a $4\pi$-kink: Bloch oscillations and current steps}
In order to quantitatively characterize a weakly dissipative quantum dynamics of a $4\pi$--kink we introduce the coordinate ("position") $x_{4\pi}$ of the $4\pi$--kink in the JJPA as: $x_{4\pi} = kd+d(4\pi - \varphi_k)/4\pi$, where $\varphi_k$ is the Josephson phase in the center of $4\pi$--kink located in the $k$-th cell. The velocity operator $\dot{\hat x}$ is defined in a standard way as
\begin{equation}
\dot{\hat x}_{4\pi}= \dv{E_{4\pi}(p)}{p},
\end{equation}
where $E_{4\pi}(p)$ is the dispersion law of the $4\pi$--kink (see, Eq. (\ref{eq:Epm})). Taking into account the Josephson relation $2eV=\hbar \dot \varphi$ one can obtain the voltage operator for the $4\pi$--kink as 
 $2e\hat V_{4\pi} = 4\pi \hbar \dot{\hat x}_{4\pi}/d$.

An applied external current bias $I(t)$ results in an additional term in the effective Hamiltonian $[4\pi/(2ed)]I(t) \hat{x}$, and weak dissipative effects are taken into account by tracing out the bath degrees of freedom \cite{dittrich1998quantum,likharev1985theory}. Summarizing that we obtain the equation describing the $4\pi$-kink dissipative quantum dynamics as following:
\begin{equation}\label{eq:dot_p}
\begin{aligned}
&\dot p_{4\pi} = \frac{4 \pi \hbar}{2ed} I(t) - \gamma \mu \dv{E(p)_{4\pi}}{p},
\end{aligned}
\end{equation}
where $\gamma$ is the phenomenological parameter (the inverse relaxation time) and $\mu = (4\pi)^2 E_J/(\omega_p d)^2$ is the effective mass. 


Next, we consider the case of dc current bias $I(t)=I$ and obtain the $I$-$V$ curve of a JJPA with  trapped $4\pi$--kink.
Substituting (\ref{eq:Epm}) in (\ref{eq:dot_p}) we rewrite the dynamic equation (\ref{eq:dot_p}) as
\begin{equation} \label{Dyn4pi}
\dot p = \frac{4\pi \hbar  I}{2e d} + \frac{ \gamma \mu  d \Delta^2 \sin(d p/ \hbar)}{\hbar \sqrt{(E_0 - E_1)^2 + 4 \Delta^2 \cos^2(p d/ (2\hbar)}}.
\end{equation}

As the dc bias current $I$ is smaller than the critical value $I_t$ the dynamic equation (\ref{Dyn4pi}) supports a steady solution $\dot p = 0$ resulting in the linear part of the current-voltage characteristics, $V \propto I$. The expression for $I_t$ is obtained explicitly as
\begin{equation} \label{It}
\begin{aligned}
&I_t = I_0 \sqrt{1 + \frac{\lambda^2}{2} + \lambda \sqrt{1 + \frac{\lambda^2}{4}}}\\
&I_0 = \frac{e d^2 \gamma \mu}{8 \Delta \pi \hbar^2}\\
&\lambda = \frac{E_1 - E_0}{\Delta}.
\end{aligned}
\end{equation}

\begin{figure}
\center\includegraphics[width = \linewidth]{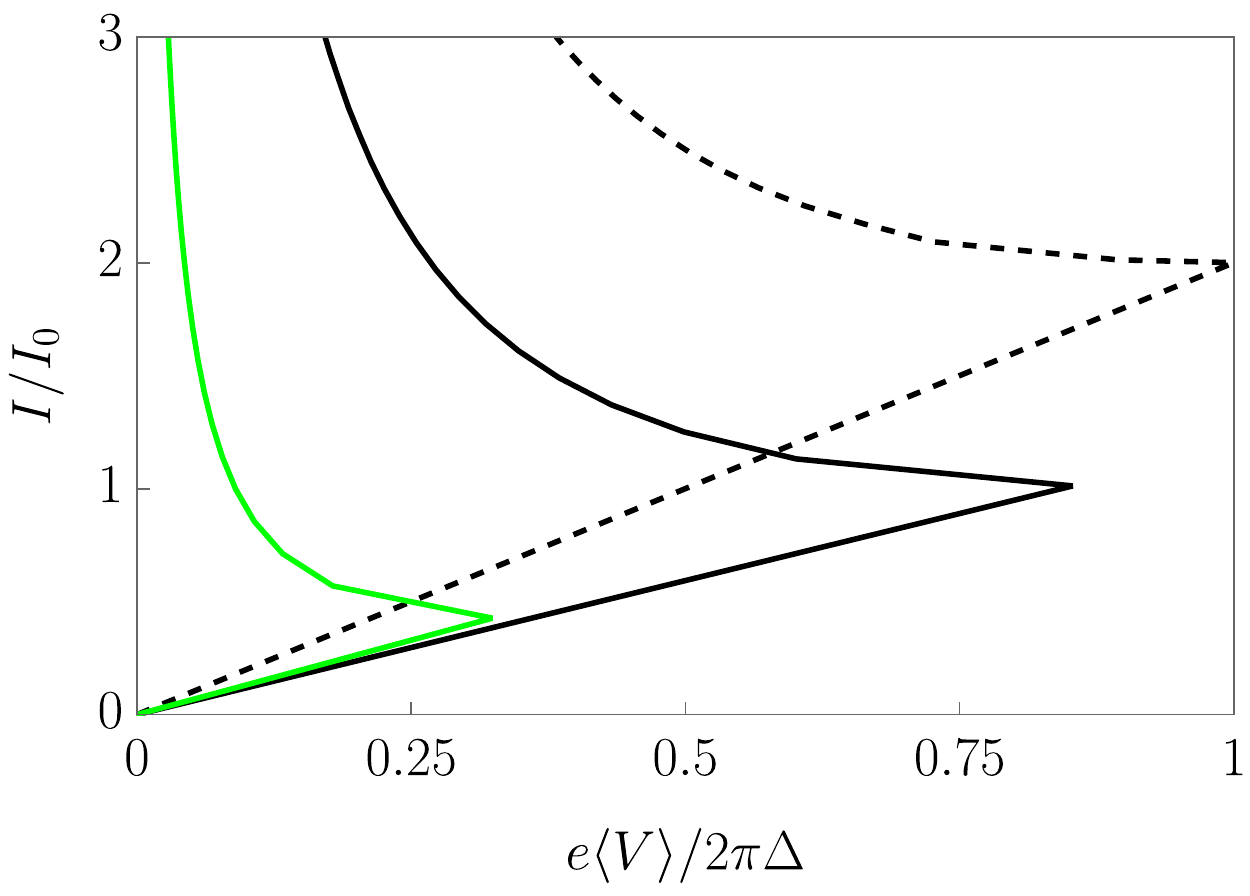}
\caption{The current-voltage characteristics of the JJPA with the trapped $4\pi$--kink. The parameters were chosen as $(E_1-E_0)/\Delta = 0$ (black solid line) and  $(E_1-E_0)/\Delta= 3$ (green solid line). For comparison the $I$-$V$ curve for two independent MFs trapped in a JJPA is shown by dashed line.}
\label{fig:IV}
\end{figure}

For $I > I_t$ the quasi-momentum $p(t)$ depends periodically on time resulting in the non-linear part of the $I$-$V$ curve. The numerical procedure to obtain the $I$-$V$ curve is following:  we fix the dc bias $I$, solve the Eq. (\ref{Dyn4pi}) numerically and calculate the time averaging of the voltage $\langle V_{4\pi}(t) \rangle$. After that we increase the dc bias current $I$ and repeat the procedure. Obtained in this way the $I$-$V$ curves are presented in Fig. \ref{fig:IV}. Thus, one can see the Bloch nose type of the current-voltage characteristics also for the $4\pi$-kink dynamics. The $I$-$V$ curve of the $4\pi$--kink deviates substantially from the one obtained for two \textit{independent} MFs. Indeed, for a fixed dc current bias the average voltage drop $\langle V_{2\times 2\pi} \rangle$ is just twice the voltage of a single MF. The current-voltage characteristics of a JJPA with trapped two independent MFs is shown in Fig. \ref{fig:IV} by dashed line. 

For $I>I_t$ the voltage $V_{4\pi}(t)$ demonstrates large amplitude periodic oscillations, i.e., Bloch oscillations. In the limit of $I \gg I_t$ the frequency of
$4\pi$--kink Bloch oscillations is $f^{Bl}_{4\pi}=I/e$. 
In the presence of both dc current $I$ and ac current with the frequency $f$ the resonance between the Bloch oscillations and an external ac current leads to the seminal current steps \cite{likharev1985theory,schon1990quantum} located at $I^{(n)}_{4\pi}=enf$. It is important to stress here that the current steps values $I^{(n)}_{4\pi}$ for the $4\pi$--kink are two times less than ones for two independent MFs, i.e. $I^{(n)}_{4\pi}=(1/2)I^{(n)}_{2\times 2\pi}$.  

\section{Conclusion}
In conclusion we present a detailed theoretical study of the quantum dynamics of two magnetic fluxons trapped in a JJPA with large kinetic inductances. In such JJPAs the characteristic size of a single MF is less than the cell size. Characterizing the Josephson phase distribution of a single MF by three consecutive Josephson phases we derive the effective potential energy of two interacting MFs. The two MFs repel each other as they occupy the same or neighboring cells. In spite of this repulsion a stable state of two merged MFs, i.e., a $4\pi$--kink, can be observed in such JJPAs. 

The quantum dynamics of $4\pi$--kink in a JJPA is determined by the energy band spectrum $E_{4\pi}(p)$ (see, Eq. (\ref{eq:Epm})). In the coherent quantum regime such spectrum leads to the quantum beats of the time-dependent return probability to observe the $4\pi$--kink on the initial point after time $t$. The amplitude and frequency of these quantum beats differ substantially from ones observed for two independently propagating MFs. 

In the presence of a weak dissipation the periodic dependence of the energy spectrum $E_{4\pi}(p)$ on the quasi-momentum $p$ results in the $4\pi$--kink Bloch oscillations with the frequency $f^{Bl}_{4\pi}$ determined by an applied dc bias current, $I$, and  the  nonlinear current-voltage characteristics with the typical "Bloch nose". The resonance between the intrinsic Bloch oscillations and an externally applied ac current provides current steps with values $I^{(n)}_{4\pi}=enf$ that two times less than ones for two independent MFs.

\textbf{Acknowledgement}
We thank Sergej Mukhin and Alexey Ustinov  for fruitful discussions. This work was partially supported by the Ministry of Science and Higher Education of the Russian Federation in the framework of the State Program (Project No 0718-2020-0025). 
\input{main.bbl}
\end{document}

%% file: main.bbl
%